\documentclass{PoS}
\usepackage{lineno}
\title{All-sky sensitivity of HAWC to Gamma-Ray Bursts}
\author{
        \speaker{Joshua Wood} $^a$ for the HAWC Collaboration$^b$ \\
        \llap{$^a$}Department of Physics, University of Maryland, College Park, Maryland, USA \\
        \llap{$^b$}For a complete author list, see 
                  \href{http://www.hawc-observatory.org/collaboration/icrc2015.php}{www.hawc-observatory.org/collaboration/icrc2015.php} \\
        Email: \email{joshwood@umd.edu}
       }
\abstract{
The High Altitude Water Cherenkov (HAWC) Observatory is a ground-based TeV gamma-ray observatory in the state of Puebla, Mexico at an altitude of 4100 m. Its 22,000 m$^2$ instrumented area, wide field of view ($\sim$2 sr), and >95\% uptime make it an ideal instrument for discovering gamma-ray burst (GRB) emission at $\sim$100 GeV. Such a discovery would provide key information about the origins of prompt GRB emission as well as constraints on extra-galactic background light (EBL) models and the violation of Lorentz invariance. We will present prospects for discovering GRB emission at $\sim$100 GeV with a simple, all-sky search algorithm using HAWC data that is most sensitive to short GRBs. The search algorithm presented here can also be used to detect other short transients with timescales and fluxes similar to short GRBs.
}

\FullConference{
  The 34th International Cosmic Ray Conference, \\
  30 July - 6 August, 2015 \\
  The Hague, The Netherlands
  }
\ShortTitle{HAWC All-Sky GRB Sensitivity}

\begin{document}

\section{Introduction}

Gamma-ray bursts (GRBs) are the most luminous events in the known universe \cite{Gehrels:2009qy}. They consist of intense gamma-ray flashes coming from cosmological distances and are roughly divided into two populations, short and long, by the duration of their gamma-ray emission \cite{Kouveliotou:1993yx}. Afterglow measurements have confirmed that long GRBs originate from the collapse of massive stars and point to the merger of compact object binaries as the progenitor for short GRBs \cite{Berger:2013jza}.

Regardless of the progenitor, the basic theory behind GRB emission is the same. It involves a central black hole powering a highly relativistic jet with gravitational energy released during the infall of surrounding matter. The jet interacts both with itself and surrounding material to form internal and external shocks where Fermi acceleration takes place, thereby generating a non-thermal spectrum \cite{Meszaros:2006rc} \cite{Piran:1999kx}.

Measurements of the highest energy gamma-rays are key to developing models of the relativistic jet inside a GRB. This is because observations of a spectral cutoff can be interpreted as estimates of the bulk Lorentz factor $\Gamma$ in the region where gamma-rays are produced \cite{Bregeon:2011bu}, providing insight into the internal GRB environment as well as the expected neutrino flux from GRBs which is sensitive to $\Gamma$ \cite{Waxman:1997ti}. Alternatively, interpreting the spectral cutoff as attenuation of GRB photons from pair-production on extra-galactic background light (EBL) results in constraints on the density of EBL over cosmological distances \cite{Gilmore:2009zb}.

In addition, the highest energy GRB photons are most sensitive to possible violations of Lorentz invariance \cite{Biesiada:2009zz}. Combining the measured arrival times of both low and high energy GRB photons yields constraints on the violation of Lorentz invariance \cite{Ackermann:2009aa}. The strongest constraints for a given GRB duration and redshift come from measuring the highest energy gamma-rays.

Since the highest energy gamma-rays carry information relevant to understanding a number of different phenomena, three major classes of high-energy gamma-ray detectors exist: satellite detectors \cite{Gehrels:1994} \cite{Atwood:2009ez} \cite{Meegan:2009qu} \cite{Gehrels:2004}, Imaging Atmospheric Cherenkov Telescopes (IACTs) \cite{Hinton:2008ka}, and Extensive Air Shower (EAS) arrays \cite{Sinnis:2009zz}. The High Altitude Water Cherenkov (HAWC) Observatory belongs to the EAS class of instruments, which combine the wide field of view ($\sim$2 sr) and high duty cycle of satellite detectors with the high sensitivity for $>$100 GeV photons characteristic of IACTs \cite{Abeysekara:2011yu}. These traits make the HAWC Observatory an ideal instrument for measuring the highest energy GRB photons.

\section{The HAWC Observatory}

The HAWC Observatory is a ground-based TeV gamma-ray observatory in the state of Puebla, Mexico at an altitude of 4100 m. Inaugurated in March 2015, this detector continuously measures the arrival time and direction of both cosmic- and gamma-ray primaries within its roughly 2 sr field of view. It is most sensitive to gamma-ray energies ranging from 100 GeV - 100 TeV and employs cuts to differentiate them from cosmic-rays, which make up the majority of the overall air shower rate \cite{Abeysekara:2013tza}.

The detector's 250 tank configuration has an overall air shower trigger rate of 10 kHz. These showers are reconstructed in near-real time and made available for analysis via a ZeroMQ socket \cite{ZMQ}, allowing for low latency searches of the overhead sky. The average latency between the time of an air shower trigger and when we can finish analyzing it is 5 seconds. This is much smaller than the $\sim$1 minute latency for checking the status of the experiment and sending an alert email. The main latency of reporting transient discoveries is therefore not the analysis itself but rather the framework used to report results.

\section{Search Method}

\begin{figure} [ht!]
  \centering
  \includegraphics[width=80mm]{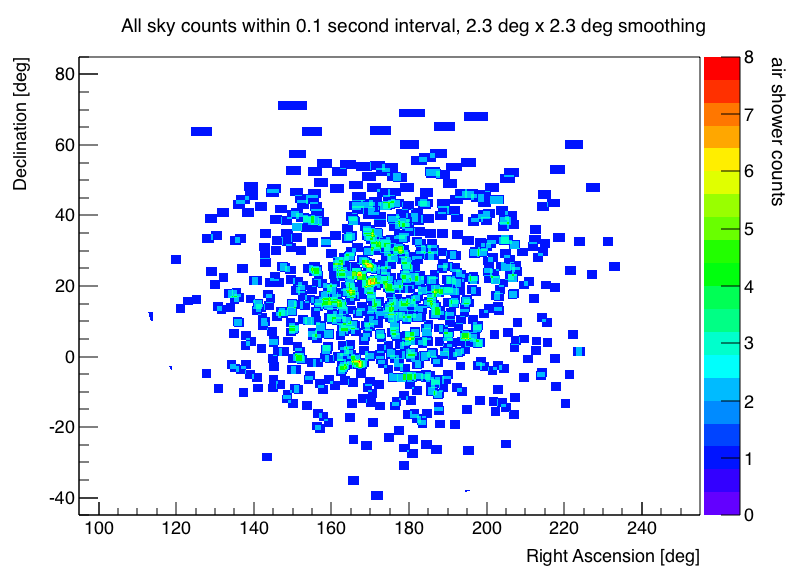}
  \caption
  {
    Sky map in equatorial coordinates of recorded air shower counts for the 0.1 second time interval. The 0.1 second time interval is chosen to limit the number of showers present in the overhead sky and produce a sparse number of events near the edges of the field of view. This illustrates our smoothing method, which uses locally square 2.3 degree $\times$ 2.3 degree spatial bins that are optimized for $\sim$100 GeV gamma-rays.
  }
  \label{fig:sigsky}
\end{figure}

Our GRB search algorithm connects to the ZeroMQ socket of reconstructed air showers and looks for GRB signals in near-real time. It does so by examining the number of air showers passing through the overhead sky within three different time intervals: 0.1, 1, and 10 seconds (see Figure \ref{fig:sigsky}). For each interval, all points within 60 degrees of detector zenith are tested against the hypothesis that the air shower count comes from the 10 kHz rate of cosmic-ray air showers. Air shower counts significantly higher than the expected background are interpreted as GRB signals.

We limit our search to zenith angles less than 60 degrees where we have the best sensitivity to $\sim$100 GeV photons \cite{Abeysekara:2011yu}. Additionally, we smooth each event with 2.3 degree $\times$ 2.3 degree square bins. This is equivalent in area to the optimal round bin predicted by the 0.8 degree angular resolution for $\sim$100 GeV photons in the final 300 tank configuration of the HAWC Observatory \cite{Abeysekara:2013tza} with a slight degradation to account for the smaller detector. We choose to use square rather than round bins because they have the advantage of being computationally efficient, allowing us to search the sky more finely.

We perform the fine spatial search by evaluating air shower counts within a 2.3 degree $\times$ 2.3 degree square bin shifted across the sky in 0.11 degree steps along the directions of right ascension and declination. This allows for a minimum overlap of $\sim$90\% between two adjacent spatial search bins. Likewise, we increment each time interval forward by 1/10th the interval duration to yield 90\% overlap in the time domain as well. This provides a significant amount of correlation between each time and spatial bin combination, allowing for fine tuning on potential GRB signals.

The 0.1 and 1 second time intervals are chosen to be representative of short GRBs. In principle, the algorithm can go down to 10$^{-4}$ seconds while remaining in real time but we are still testing timescales $<$ 0.1 seconds. The maximum interval of 10 seconds exists because it is on the order of the time over which we can treat the sky as static given the 0.11 degree separation between spatial search bins. We are working to extend the algorithm to longer intervals by accounting for the Earth's rotation.

\section{Background Estimation}

\begin{figure} [ht!]
  \centering
  \includegraphics[width=110mm]{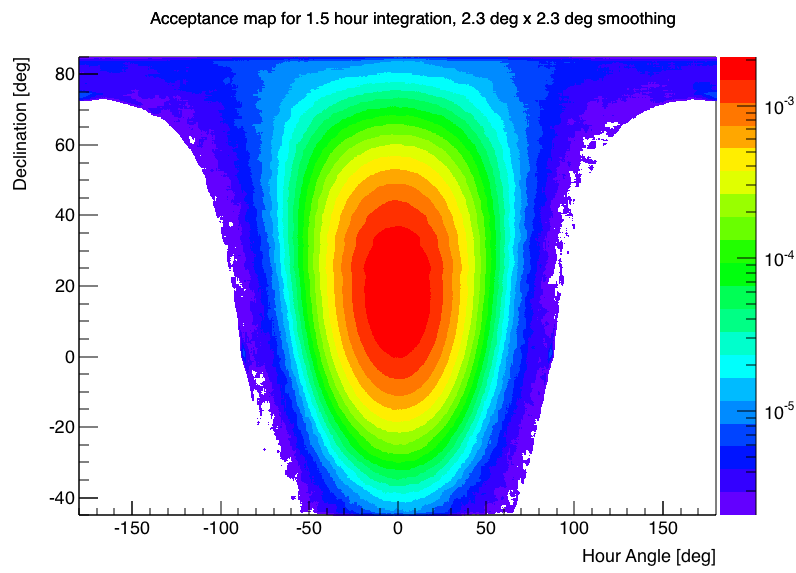}
  \caption
  {
    Background integrated for 1.5 hours and normalized to one.
    The integration time scale is set by the minimum time needed to accurately estimate background in spatial bins far from zenith.
  }
  \label{fig:backmap}
\end{figure}

At the moment, we do not have an optimized set of cuts to discriminate between cosmic- and gamma-ray air showers in this analysis. The entire 10 kHz trigger rate of mostly cosmic-ray air showers therefore makes up the background for this search. This rate changes slightly with atmospheric density fluctuations over the course of each day, however, the air shower arrival distribution over the HAWC Observatory's field of view remains fairly constant \cite{Abeysekara:2013tza}. We therefore estimate the background in each search bin by integrating air showers in a map of declination versus hour angle over a 1.5 hour period. We normalize this map to one and smooth it into the 2.3 degree $\times$ 2.3 degree square bins used in the spatial search to obtain the fraction of total showers entering each search bin (Figure \ref{fig:backmap}).

\begin{figure} [ht!]
  \centering
  \includegraphics[width=110mm]{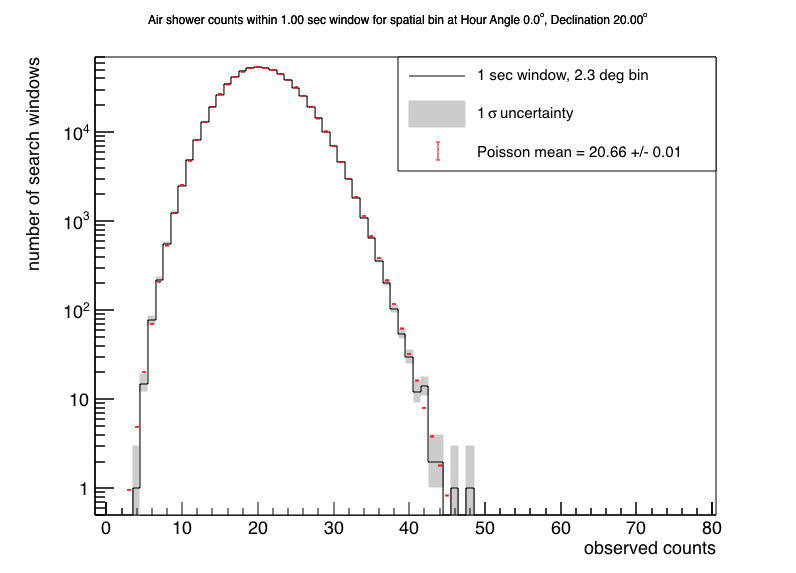}
  \caption
  {
    Histogram of observed counts near detector zenith over the course of one day for the 1 second interval search. The data follow what one expects from a Poisson distribution with the same mean. The errors bars for the Poisson expectation are smaller than can be seen using this vertical scale.
  }
  \label{fig:poisson}
\end{figure}

Multiplying a specific point in Figure \ref{fig:backmap} by the duration of a given search interval and the current air shower rate measured at the center of this interval gives the expected number of background counts for that location. For example, the 1 second duration search yields approximately (1 second $\times \sim$10 kHz $\times \sim$2$\times 10^{-3}$) $=$ 20 expected counts at a declination of 20 degrees and an hour angle of 0 degrees. The observed number of counts for this expectation follow a Poisson distribution (Figure \ref{fig:poisson}) so we quantify the significance of upward fluctuations using Poisson statistics.

\section{Speedup Methods}

\begin{figure} [hb!]
  \centering
  \includegraphics[width=130mm]{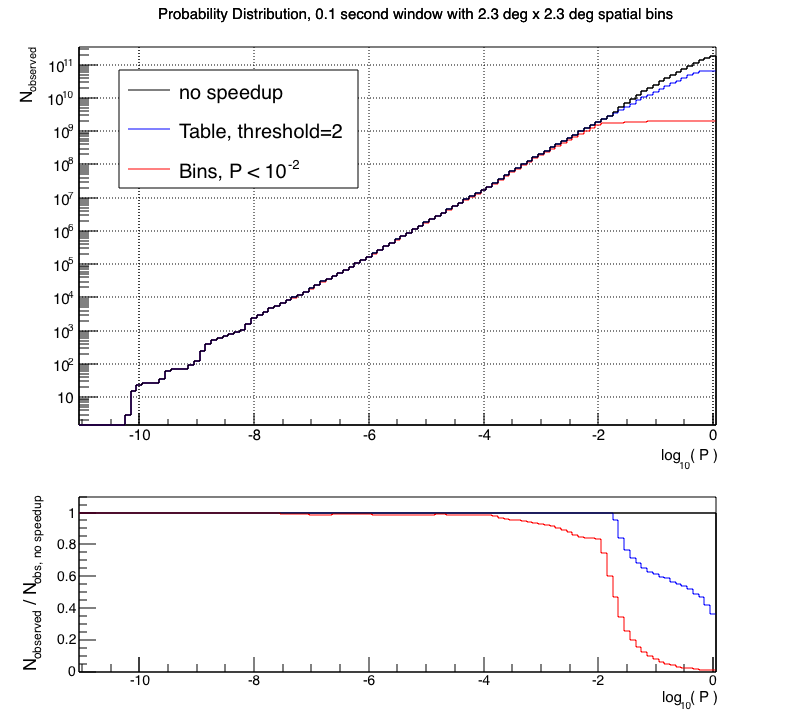}
  \caption
  {
    Figure showing preservation of probability distribution after speedup measures. The black line is the full probability distribution for the 0.1 sec duration search over 2 hours of data without speedup techniques applied. The blue line is the probability distribution for skipping bins with $<$ 2 counts. The red line is the probability distribution from a coarse/fine search with a transition at P $<$ $10^{-2}$. The equivalence of all three curves below P $<$ $10^{-8}$ means the Table and Bins speedup methods preserve both the rarest events and the total number of trials from the full search.
  }
  \label{fig:speedup_pdist}
\end{figure}

We employ three main speedup techniques in order to ensure our search algorithm can keep up with the 10 kHz air shower rate. The first method only applies to search durations < 0.1 seconds for which most bins in the sky will be empty. In this case, we choose not to evaluate spatial bins containing $<$ 2 counts because they cannot provide upward fluctuations significant enough to claim as evidence for GRB signals after accounting for trials. We refer to this method as the Table method since information about bins containing $\ge$ 2 counts are stored in a lookup table. Figure \ref{fig:speedup_pdist} shows that the Table method preserves the overall probability distribution below P $=$ $10^{-8}$, where we expect signal.

The second speedup method applies only to window durations > 0.1 seconds, where we expect a large fraction of sky bins to be filled. In this case, we group 3 $\times$ 3 blocks of adjacent spatial search bins and first perform a coarse search over the central bin in each group. We then only perform a fine search in the remaining 8 bins when the central bin has a probability lower than $10^{-2}$. Again, this preserves all the unlikely events while reducing the sky map evaluation time by a factor of $\sim$10. We refer to this as the Bins method because it is a full search over all spatial bins in the current field of view.

The third and final speedup method is the pre-calculation of Poisson probabilities in a lookup table, which we use for all search durations. To do this, we discretely bin the expected background counts from the smallest possible expectation, obtained by multiplying the smallest non-zero count rate within our 1.5 hour background integration, 1/5400 Hz, by the smallest time window of 0.1 seconds, to twice the largest count measured at zenith in the 10 second search window. We choose a fixed, logarithmic spacing between each binned background count which is finer than 10\% the $\sqrt{N}$ uncertainty at the largest expected background count. We then calculate the compliment of the cumulative Poisson probability for observing between 0 and 5000 counts in each background bin to create lists of cumulative probabilities indexed by background bin and observed count. The maximum of 5000 observed counts is much greater than the largest background count plus the $\sim$100 GRB signal photons expected from the HAWC Observatory's effective area being 100 times larger for 100 GeV photons than that of the Fermi satellite \cite{Atwood:2009ez} \cite{Abeysekara:2013tza}.

We obtain probabilities during the search by converting expected background counts to bins within the pre-calculated probability table and looking up the entry for a given observed count. Recording the probability result from every searched time window and spatial point over a 2 hour interval yields the black curve shown in Figure \ref{fig:speedup_pdist}. This curve matches a line with slope 1, proving that our background binning is fine enough to accurately estimate all necessary background counts.

\section{Sensitivity}

The benefit of searching the whole sky is the possibility of finding signals in portions of the HAWC Observatory's field of view where satellites are not overhead. These signals would otherwise be missed if we restricted our search to locations of GRBs reported by external notices from the GRB Coordinates Network (GCN). However, there is a sensitivity cost associated with the trials taken to continuously search the full sky.

For example, we incur 8.64$\times 10^5$ temporal trials when shifting our 1 second search window in time by 10\% its width over the course of 1 day. We also gain another 1.33$\times 10^6$ spatial trials when searching the field of view using 0.11 degree steps in right ascension and declination over all areas within 60 degrees of zenith for each temporal trial. This yields a total of 1.15$\times 10^{12}$ trials taken each day by our 1 second search window and requires a signal increase from 27 photons to 52 photons for a 5$\sigma$ discovery at our current expected background of 20 counts near detector zenith if we treat each trial as independent.

The overlap between each spatial and time bin used in our search technique means our trials are not independent. We are still working on an a method to compute the effective number of trials taken in our search, but the result will be smaller than the total number of trials and therefore requires a smaller flux increase for a 5$\sigma$ discovery. Regardless, the roughly 2$\times$ worse sensitivity of our full sky search compared to the single trial sensitivity presented in reference \cite{Abeysekara:2011yu} when treating trials independently equates to a differential sensitivity of E$^2$ dN/dE $\approx$ 4$\times 10^{-6}$ erg cm$^{-2}$ s$^{-1}$ at 10 GeV assuming a spectrum of the type dN/dE $\propto$ E$^{-2}$ with a high-energy cutoff at 100 GeV for the current trigger of nHit $>\sim$30. This makes the detection of bursts like GRB 090510 or GRB 090902b possible if the high-energy cutoff is indeed above 100 GeV.

\section{Conclusions}

We show that our simple, all sky search method has a differential sensitivity of about 4$\times 10^{-6}$ erg cm$^{-2}$ s$^{-1}$ at 10 GeV assuming a spectrum of the type dN/dE $\propto$ E$^{-2}$ with a high-energy cutoff at 100 GeV. This search is performed in near-real time and does not rely on external information from other experiments. The largest latency in our setup is checking the status of the experiment and sending the notice of a detection to the community.

In principle, our search method is not tuned to GRB signals alone. It is sensitive to any transient with a timescale and photon emission spectrum similar to those of GRBs, such as evaporating primordial black holes (PBHs) \cite{Abdo:2014apa}. PBHs are easily distinguished from GRB signals using the abrupt cutoff in their light curves, which corresponds to the end of evaporation \cite{Abdo:2014apa}.

\section*{Acknowledgments}
\footnotesize{
  We acknowledge the support from: the US National Science Foundation (NSF);
  the US Department of Energy Office of High-Energy Physics;
  the Laboratory Directed Research and Development (LDRD) program of
  Los Alamos National Laboratory; Consejo Nacional de Ciencia y Tecnolog\'{\i}a (CONACyT),
  Mexico (grants 260378, 232656, 55155, 105666, 122331, 132197, 167281, 167733);
  Red de F\'{\i}sica de Altas Energ\'{\i}as, Mexico;
  DGAPA-UNAM (grants IG100414-3, IN108713,  IN121309, IN115409, IN111315);
  VIEP-BUAP (grant 161-EXC-2011);
  the University of Wisconsin Alumni Research Foundation;
  the Institute of Geophysics, Planetary Physics, and Signatures at Los Alamos National Laboratory;
  the Luc Binette Foundation UNAM Postdoctoral Fellowship program.
}

\providecommand{\href}[2]{#2}\begingroup\raggedright\endgroup

\end{document}